\documentstyle[prl,aps,psfig,multicol]{revtex}

\begin{document}
\title{Small polaron formation in many-particle states of the
Hubbard- Holstein model: The one-dimensional case}        
       
\author{Massimo Capone}
\address{Istituto Nazionale di Fisica della Materia and
International School for Advanced Studies (SISSA-ISAS),\\
Via Beirut 2-4, Trieste, Italy 34013}
\author{Marco Grilli}
\address{Istituto Nazionale di Fisica della Materia and
Dipartimento di Fisica, Universit\`a di Roma ``La Sapienza'',\\
Piazzale A. Moro 2, Roma, Italy 00185}
\author{Walter Stephan}
\address{Department of Physics and Astronomy, University of
Manitoba, Winnipeg, Manitoba, Canada R3T 2N2}
\date{\today}
\maketitle

\begin{abstract}
We investigate polaron formation in a many-electron system
in the presence of a local repulsion sufficiently strong to prevent
local-bipolaron formation. Specifically, we consider a Hubbard-Holstein 
model of interacting electrons coupled to dispersionless phonons
of frequency $\omega_0$. Numerically solving the model in
a small one-dimensional cluster, we find that in the nearly
adiabatic case $\omega_0 < t$, the necessary and sufficient 
condition for the polaronic regime to occur 
is that the energy gain
in the atomic (i.e., extremely localized) regime  ${\cal E}_{pol}$ 
overcomes the
energy of the purely electronic system $ {\cal E}_{el}$. In the
antiadiabatic case, $\omega_0 > t$, polaron formation
is instead driven by the condition
of a large ionic displacement $g/\omega_0 >1$ ($g$ being the
electron-phonon coupling).
Dynamical properties of the model in the weak and moderately
strong coupling regimes are also analyzed.

\end{abstract}

\begin{multicols}{2}
\section {INTRODUCTION}
There is increasing evidence that polarons are formed in the lightly
doped insulating phase of the high temperature superconducting
cuprates \cite{polaronexper,polaronexper2} and in 
the high-temperature paramagnetic phase in doped manganites \cite{millis}. 
On the theoretical side, simple polaronic models like the Holstein and
the Su-Schrieffer-Heeger models have attracted
intensive studies in the last few years.
However, whereas the cases of one or two
polarons have been carefully investigated both numerically
\cite{deraedt1a,rt,dmr,wellein,CSG,CCG,DMRG} 
and analytically \cite{ciuchi,CDFF},
the case of many polaronic carriers is still incompletely understood.
The relevance of electronic correlations in the materials mentioned 
above provides a strong motivation to study the combined
effect of the electron-phonon coupling which leads to polaronic
features and the strong electron-electron interaction.
The main goal of the present work is to numerically investigate
strongly correlated many-particle systems interacting with 
phonons in order both to identify 
a simple criterion for
the formation of a polaronic state and to characterize the
static and dynamical properties of such a state.
We  will only be concerned with  the Hubbard-Holstein
and the $tJ$-Holstein models, 
as minimal systems with local electron-electron (e-e) and 
electron-phonon (e-ph) interactions. In order to focus on the physics
of unbound polarons (as opposed to bipolarons), we will always
consider the limit of strong local e-e repulsion to
prevent the formation of a local bipolaron  \cite{notenobipol}. 
The short-range character of the 
bare interactions considered in this model greatly simplifies the
numerical analysis on finite clusters. Furthermore, the 
strong coupling limit of the e-ph interaction
gives rise to small (single-site) polarons, which are individually
rather well understood, thus providing a simple limit of the model.
Since we focus on
metallic states, we purposely avoid specific conditions (like, e.g., 
quarter-filling) leading to ordered insulating states. 
Finally we confine our analysis to the one-dimensional case.
This choice is an inescapable consequence of the smallness of our
numerical clusters, but also allows for useful comparisons with
the wealth of well-established physical results
available in $d=1$. 

The model reads
\begin{eqnarray}
{\cal H}& = & -t\sum_{\langle ij\rangle, \sigma}
\left( c^\dagger_{i,\sigma}c_{j,\sigma} + h.c. \right) +
U \sum_{i}c^\dagger_{i,\uparrow}
c_{i,\uparrow} c^\dagger_{i,\downarrow}c_{i,\downarrow} \nonumber \\
&+& g\sum_{i,\sigma} 
\left[ c^\dagger_{i,\sigma}c_{i,\sigma}-
\langle c^\dagger_{i,\sigma}c_{i,\sigma}\rangle \right]
 \left( a_i+a^\dagger_i \right)
+\omega_0 \sum_i a^\dagger_i a_i.
\label{HH}
\end{eqnarray}
We use units such that the 
lattice spacing $a=1$ and also $\hbar=c=1$. To make our analysis
more complete, we also investigate the $tJ$-Holstein model, where
the $U$ term in (\ref{HH}) is replaced by an Heisenberg
interaction ${\cal H}_J= J \sum_{\langle ij\rangle}
({\bf S}_i \cdot {\bf S}_j -{1 \over 4} n_i n_j)$, 
with the additional constraint of no double
occupancy $\sum_\sigma c^\dagger_{i,\sigma}c_{i,\sigma} \le 1$.

\section{THE SMALL POLARON CRITERION}

A polaronic state can be characterized as a bound state of 
electrons and phonons, in which the electronic motion
is accompanied by a significant lattice displacement.
The strong coupling between the electron and the lattice
strongly suppresses the electronic mobility. 
The carriers are therefore self-trapped in the potential well
that they generated.
The polaronic state is therefore characterized by two
conditions: {\it{(a)}} the energy gain associated with the self-trapping
must be larger than the loss of kinetic energy, and {\it{(b)}} the local
lattice displacement must be sizeable, in 
order to significantly reduce the electronic hopping amplitude.

In  previous analyses in the single-particle case
of the Holstein model\cite{CSG,CDFF,CCG}, 
emphasis was put on the difference in the 
criteria for polaron formation in the adiabatic ($\omega_0<t$ ) 
or antiadiabatic ($\omega_0>t$) regimes.
In the adiabatic regime the crossover to polarons is dictated by
the condition {\it{(a)}}, since in this limit, as soon as a bound
state is energetically favorable, the electron mobility is
automatically reduced due to the large mass of the lattice.
In the single particle case the energy of the
strong-coupling polaronic state is ${\cal E}_{pol} = -g^2/\omega_0$,
while the free electron energy  ${\cal E}_{el} = -2dt$. 
Then the condition for 
single polaron formation in the nearly adiabatic
($\omega_0/t < 1$) case reads (see, e.g. Ref.\onlinecite{CSG}). 
\begin{equation}
\label{1particle}
\lambda \equiv g^2/(2dt\omega_0) >1.
\end{equation}
In the antiadiabatic regime the 
crossover for the single-particle case is instead given
by the condition {\it{(b)}}, which in the Holstein model is
expressed as $\alpha \equiv g/\omega_0>1$.

In the presence of many carriers and e-e correlations
the electronic energies and the e-ph coupling are affected, 
so that the criterion for polaron formation 
in the adiabatic regime (which involves a condition on the energies)
will be different in the many particle case
with respect to the single particle one. 
On the other hand condition {\it{(b)}} ruling polaron formation
in the antiadiabatic regime
is not substantially changed by increasing the number of particles.
In particular the slow dynamics of the electronic degrees of
freedom when $\omega_0 \gg t$ strongly suppresses electronic
screening processes. Therefore in the antiadiabatic regime
the single-particle condition $\alpha = g/\omega_0 > 1$ 
of large ion displacement is
expected to hold also in the many particle case.  

For these reasons we mostly investigate
here  the physically relevant case of the adiabatic regime where
the phonon frequency $\omega_0$ is smaller than the 
electronic energy scale $t$ (the typical electronic energy scale
stays $t$ even in the presence of correlations).
 The same analysis has also been applied 
to the antiadiabatic case, where it is confirmed that
the single-particle condition
$\alpha > 1$ holds in the many-particle case as well
for all the considered values of parameters and
fillings. Therefore we will only present results for the adiabatic regime.
In this case the energetic balance condition rules the crossover. 
A polaronic bound state can therefore be realized only if the energy of
such a state ${\cal E}_{pol}$ is lower than the energy 
of the electrons in the absence of electron-phonon interaction
${\cal E}_{el}$.

It is quite natural to generalize the condition (\ref{1particle}) 
to the many-particle case by comparing 
 the energy of the strongly coupled e-ph system,
${\cal E}_{pol}$, where the polarons are strongly
localized and the kinetic energy is negligible, with
the energy of the purely electronic system  
(i.e. of the simple Hubbard or $tJ$ models) .
In other words, we identify the value of the coupling for which the polaronic
crossover occurs, $g=g_c$, with the e-ph coupling above which the
energy of the strongly polaronic state is lower than the
energy of the purely electronic state.
In this framework it is useful to
generalize the definition of $\lambda$ by defining the  quantity 
\begin{equation}
\label{lambdatilde}
\tilde\lambda \equiv {\cal E}_{pol}(g)/{\cal E}_{el}
\end{equation}
that reduces to $\lambda$ for a single particle, and show that 
the criterion for small polaron formation in the general case 
of finite densities and in the presence of $e$-$e$ correlation is
$\tilde\lambda > 1$.
Of course this 
criterion can only be significant provided the crossover between
the purely electronic and the polaronic regimes is
sufficiently sharp, as is indeed the case in the adiabatic regime\cite{CCG}. 

The criterion $\tilde\lambda \equiv {\cal E}_{pol}(g)/{\cal E}_{el}>1$
would be of little use if simple
expressions in terms of the bare parameters or at least 
simple estimates of ${\cal E}_{pol}(g)$ and ${\cal E}_{el}$
were not available. Therefore, we now turn to the 
explicit evaluation of ${\cal E}_{pol}(g)$ and ${\cal E}_{el}$.
Unfortunately, 
while some simple arguments allow an easy estimation of the
energy in the strongly polaronic regime, the knowledge of the
purely electronic energy (i.e. of the Hubbard model) is a much
more difficult task. However, we point out that our goal is {\it not}
to investigate the Hubbard model by  itself, but rather to
analyze the many-body effects of interaction on polaron formation,
once a full knowledge of the purely electronic system (no matter how
complicated) is assumed.

In order to evaluate the strong coupling energy  ${\cal E}_{pol}(g)$
we start from the atomic
limit of zero hopping $t=0$, in which the Hubbard-Holstein model
can be exactly solved. In this 
case a Lang-Firsov canonical transformation,
 allows the elimination of the linear e-ph coupling
by introducing a shift $\alpha\equiv g/\omega_0$ of
the ionic equilibrium position. 
An effective 
non-retarded attraction between the particles arises
$U_{attr}=-2g^2/\omega_0$, which must be compensated by
a larger repulsion $U$ to avoid (local) bipolaron formation. 
In this atomic limit each site is independent
and each electron individually shifts the ionic site on which
it resides forming a strong-coupling polaron of energy 
${\cal E}_{1-pol}=-g^2/\omega_0$. An opposite shift, but with
identical energy gain, is induced by holes, so that,
for $N_s$ sites at average filling $n$ the total
energy is ${\cal E}_{pol}^{t=0}=-N_sn(1-n)g^2/\omega_0$.
As soon as a finite hopping is introduced, a coherent kinetic 
energy contribution arises, which, even in the absence of e-e 
interaction is exponentially small [$t^*=t\exp(-\alpha^2)$]
and can safely be neglected in our estimate of the energy.
However, more relevant  additional corrections arise, which
are due to incoherent virtual hopping processes of the electrons
to neighboring sites. These processes naturally depend 
on the occupancy of the neighboring sites. If the site
is unoccupied, the electron can hop onto it without paying any
electronic repulsion $U$. However, this hopping occurs
instantaneously without allowing the lattice time
to relax. Therefore an intermediate virtual state 
of higher energy proportional to $g^2/\omega_0 \gg t$ is reached, before
a second hop restores the initial configuration.
In the limit of very strong e-ph coupling and accounting for the
probability for a singly occupied site to have at least
one empty site nearby, the virtual double-hopping
process lowers the total energy by an amount of the order\cite{nota1sulambda}
\begin{equation}
{\cal E}_{1/\lambda} \approx -2 N_sn(1-n)t^2/(g^2/\omega_0) 
\label{epola}
\end{equation}
On the other hand a second incoherent correction to the polaronic
energy occurs when the neighboring site is already occupied
by an electron with opposite spin (for parallel spins this process
is forbidden by the Pauli principle). In this case a virtual state
is reached where an e-e interaction is also present
giving rise to a superexchange coupling.  In the case of 
finite e-e repulsion $U$, this term provides an additional 
incoherent contribution to the energy of the polaronic state. 
In a recent work \cite{SCGC} we showed that the e-ph coupling dresses 
the effective magnetic coupling $J$ leading to an increase of its
value.  However, for a Holstein-like
e-ph coupling no corrections to the purely electronic $J$
arise up to order $g^2/U$.  For the sake of simplicity
we therefore estimate this contribution ${\cal E}^{J}_{el}$
by using the value for $g=0$ calculated 
by exact diagonalization.

As a result, the energy of the polaronic state can be estimated as
\begin{equation}
{\cal E}_{pol}\approx -N_sn(1-n)g^2/\omega_0 +{\cal E}_{1/\lambda}+
{\cal E}_{el}^J.
\label{Epolar}
\end{equation}
It is worth noting that in the above expression the
bare e-ph coupling and phonon frequency appear. This is because
in the strong-coupling regime, where the coherent motion of the
polarons is suppressed, no screening occurs due to the
very massive carriers. This situation is different from the case
of a Fermi liquid, where the presence of the e-e
interaction leads to a dressing of the e-ph coupling
\cite{GC} and is complementary to the case of an interacting electron system
near the Mott-Hubbard transition (or near the Luther-Emery point
for one-dimensional systems), where Umklapp processes render
the e-ph interactions ineffective: If a system is localized
by the e-e interaction, this latter screens out the effects of
the e-ph coupling and, conversely,  if a system  is localized
by strong e-ph interactions (as in the present
case), the e-e interactions have no effect
on the phonon parameters.

The evaluation of the purely electronic energy ${\cal E}_{el}$
of the Hubbard model is not possible in general 
unless one resorts to numerical methods. However,
for the sake of simplicity, we consider a one-dimensional system
for which we can evaluate the energy in a simple way
using well-known results.
In the $U=\infty$ limit an exact mapping exists between the
Hubbard model and a system of free spinless fermions \cite{ogatashiba},
and in the more general case of finite $U$ 
the Bethe ansatz solution of the Hubbard model provides the exact
results (see, e.g., Ref.\onlinecite{shiba}). 
To make our calculation similar for the Hubbard and the $tJ$ model,
the alternative we chose is to directly carry out a
numerical evaluation of the ground state energy of the purely electronic
models on a one-dimensional cluster as a function of $U$ or $J$ and $n$.

We are now in a position to check the validity of the
criterion $\tilde\lambda_c \equiv {\cal E}_{pol}(g_c)/{\cal E}_{el} \sim 1$
by means of exact diagonalization of finite clusters.                 
Due to the infinite set of accessible phonon states
on each site, we need to truncate the phonon Hilbert space by 
allowing for a finite maximum number $N_{ph}^{Max}$ 
of phonons on each site.
The crossover between quasi-free electrons and a polaronic state
is signaled by rather abrupt changes in the behavior of most
physical quantities; many quantities can be used to identify
the crossover coupling, some examples of which are the electronic kinetic
energy or effective mass, the electron-lattice correlation function 
or the average number of phonons per site, and so on. 
We identify the crossover value  $g=g_c$ separating the
two regimes with the value of $g$ for which
the slope of the average number of phonons per site 
$\langle n_{ph}(g) \rangle$ is maximum\cite{notacriterio}. 
We
always limited the calculation to values of $g$ such that
$\langle n_{ph}(g) \rangle < N_{ph}^{Max}$ and checked that 
the result was well converged by changing $ N_{ph}^{Max}$.

We performed calculation for various fillings in 5 and 6 sites clusers
for $\omega_0/t =0.2$. In table I we report, besides the electronic
energy and the magnetic energy, the crossover values of $\lambda$ and
$\tilde\lambda$
obtained using the numerically calculated values of $g_c$.  

As far as $\tilde\lambda$ is concerned,  we report results obtained both
neglecting the incoherent $1/\lambda$ contribution 
${\cal E}_{1/\lambda}$ 
(sixth column,$\tilde\lambda_c$) and considering such contribution
(seventh column,$\tilde\lambda_c'$)

Note that for $U=\infty$ an additional
symmetry appears around quarter-filling ($n=1/2$), making
the filling $n=1/5$ equivalent to $n=4/5$ and $n=2/5$
equivalent to $n=3/5$.

It is evident from Table I that the criterion
$\tilde\lambda_c \approx 1$ is quite well satisfied, even neglecting
the incoherent term, whereas
the values of $\lambda_c$ are significantly different for different
parameters. This result is
particularly remarkable since the energies  ${\cal E}_{pol}(g_c)$ and 
${\cal E}_{el}$ entering
in the numerator and the denominator of $\tilde\lambda$ respectively, vary
significantly with filling and Coulomb repulsion $U$ (or $J$).
Various observations are in order. First of all
it is apparent that the calculation of $\tilde\lambda_c$ 
without including the incoherent contribution ${\cal E}_{1/\lambda}$
to ${\cal E}_{pol}$ gives quite homogeneous values 
of $\tilde\lambda_c \sim 1$ (within 10\%)
for systems with different values of $U$ or $J$,
but similar filling. Larger discrepancies (but always
smaller than 20\%) exist between different classes of fillings.
We notice that the values of $\tilde\lambda_c$ for finite $U$ or $J$
are slightly underestimated because we estimated the magnetic
energy in ${\cal{E}}_{pol}$ taking the purely electronic
contribution, which is known to be lower than in the presence of
e-ph coupling \cite{SCGC}. This underestimation is proportionally
more crude for larger $J$ or smaller $U$ values.
Correcting for this effect should produce
an even more uniform value for  $\tilde\lambda_c$.
The magnetic energy in the strongly polaronic regime is also different
from ${\cal{E}}_{el}^J$ because the reduced itinerancy of the
carriers obviously affects the spin correlations. Therefore
a further refinement of the estimate of ${\cal{E}}_{pol}$ could
be obtained by calculating  the magnetic energy in the polaronic state
taking ${\cal{E}}_{el}^J$ from the
purely electronic model with $t=0$. Also in this case a larger 
magnetic coutribution would be found (when the carriers are localized 
the magnetic correlations are obviously stronger) and this would go in the
direction of rendering the values for $\tilde \lambda_c $ more uniform.

An additional important ingredient in the estimate of $\tilde\lambda_c$
turns out to be the contribution ${\cal E}_{1/\lambda}$
of the incoherent hopping
processes to ${\cal E}_{pol}$ [see Eq. (\ref{Epolar})].
The estimated filling dependent contribution (\ref{epola})
leads to a sizable modification of the criterion
from $\tilde\lambda_c \sim 1$ to $\tilde\lambda_c' \sim 1.2$. 
However, despite
this purely quantitative change, which was not considered in our
previous single-polaron work\cite{CSG}, and which arises from a
simple refinement of the estimate of the polaron energy, 
the physical meaning of the criterion stays the same.
As a result of this latter filling-dependent improvement
in the estimate of the polaron energy, the crossover values
$\tilde\lambda_c$'s reported in the last column of Table I turn out
to be more homogeneous with varying filling within
each class of interaction values and the criterion $\tilde\lambda_c'
\approx 1.2$ seems to be generically satisfied within a few percent.

\section{DYNAMICAL PROPERTIES}
In the previous section, we characterized the crossover from
weak to strong coupling regimes through the analysis of
static properties, namely the average number of phonons.
In this section we consider the effects of this
polaron-formation crossover on dynamical properties
such as the spectral densities
$ A(\omega)$ and the dynamical conductivity $\sigma (\omega)$.
Here we report on 
this investigation for the specific but typical case
of the Hubbard-Holstein model with $U=10t$. 

\subsection{Spectral Function}
In this section we study the spectral density associated with the 
injection of an electron (inverse photoemission, IPE)

\begin{equation}
A^{+}_{\sigma}(\omega) = {1\over N}\sum_{k,n} \vert
\langle\phi_n^{M+1}\vert c^{\dagger}_{k\sigma}\vert\phi_0^{M}
\rangle\vert^2\delta(\omega-(E_n^{M+1}-E_0^{M})),
\end{equation}
and the corresponding quantity for the emission of an electron 
(photoemission, PE)
\begin{equation}
A^{-}_{\sigma}(\omega) = {1\over N}\sum_{k,n} \vert
\langle\phi_n^{M-1}\vert c_{k\sigma}\vert\phi_0^{M}
\rangle\vert^2\delta(\omega+(E_n^{M-1}-E_0^{M})).
\end{equation}

\begin{figure}
\centerline{\psfig{bbllx=600pt,bblly=100pt,bburx=60pt,bbury=700pt,%
figure=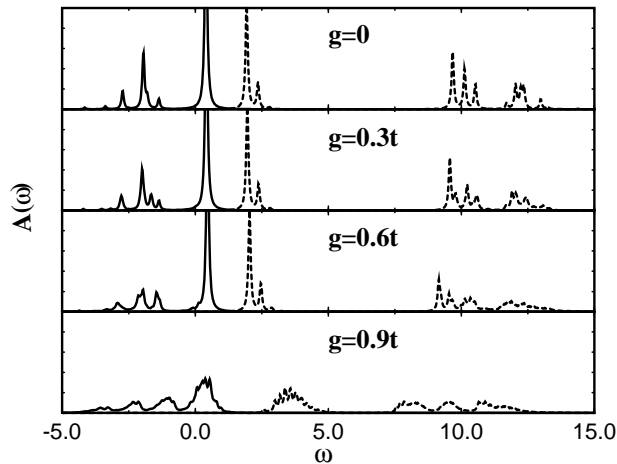,width=70mm,angle=-90}}
\caption{Spectral density (solid line for PE, dashed line for IPE)
for $U=10t$ and $\omega_0/t =0.4$, for various values of $g$ and $n=4/5$.}  
\label{akw}
\end{figure}

In Fig.\ref{akw} we show the PE (solid line) and IPE (dashed line)
spectral densities for a 5-site cluster and four electrons with
$U=10t$, for different values of $g$ ranging from $g=0$ to $g=0.9t$.

A recent $k$-resolved spectral analysis
in the two-dimensional $tJ$ model with two holes\cite{bauml},
 shows a markedly different behavior between
the incoherent (local) excitations at high energies,
not substantially affected by e-ph coupling, and 
the dispersive ``quasiparticle'' states near the Fermi level,
which become very massive with increasing $g$ \cite{notacohe}.
On the contrary
this behavior is not present in our one-dimensional 
cluster, where the fact that electrons are composites
of holons and spinons leads to all
electronic excitations in the upper and lower Hubbard
bands being incoherent.
These are only
marginally affected by the e-ph interaction and mainly
display a broadening, which is particularly evident
in the upper Hubbard band. A closer inspection also
allows one to detect the effects of multiphonon excitations dressing 
the purely  electronic states by forming ``shoulders'' or ``combs''
with excitation energies spaced by $\omega_0$. 

The presence of a phonon-induced retarded local attraction also
reduces the local instantaneous Hubbard repulsion U. 
For the moderate $g=0.3t$ and $g=0.6t$, this
attractive effect is not large enough
 ($\lesssim 2g^2/\omega_0 \lesssim 2$) to be clearly visible
in a marked reduction of the gap between the lower and upper 
Hubbard bands. Rather a small leaking of spectral weight at
 the edges of the gap results from the much more evident
broadening mentioned above. 
The effective phonon-mediated attraction
increases the amount of doubly occupied sites in the ground state
thus modifying the relative 
weight of the spectral densities related to the injection 
(inverse photoemission, IPE) or to the
emission (photoemission, PE) of electrons, 
represented by the dashed and solid
lines in Fig. \ref{akw} respectively. For the large 
value of $U=10t$ and for $g=0.3t$ and $g=0.6t$, which are 
substantially smaller than
those leading to bipolaron formation, the amount of doubly
occupied sites in the ground state is always small and its 
small changes do not significantly affect the relative 
weight of the photoemission and inverse photoemission
spectral features.
As a consequence one observes an approximate, but quite 
well-satisfied conservation of the spectral weight
in the  PE and IPE parts of the spectrum {\it separately}
for moderate values of $g$.
As the e-ph coupling is further increased to $g=0.9t$
the energy gap between the upper and the lower Hubbard band
is substantially reduced and the weight conservation
no longer holds separately for the PE and IPE parts of the spectrum.

On the other hand, for the largest values of $g$ we observe 
a substantial transfer of spectral weight from the upper Hubbard 
band, which is broadened and lowers its overall intensity, to the
lower Hubbard band. This latter acquires weight essentially because
the possibility of exciting phonons allows more states to be reached
by injecting one electron in the system by conserving the momentum
and without increasing the number of doubly occupied sites
in the final state. 

\subsection{Optical Conductivity}

The real part of the conductivity for a one-dimensional 
tight-binding model at zero temperature may be expressed in 
terms of the Kubo formula
\begin{equation}
\sigma(\omega) = D\delta(\omega) + 
\sum_{n\ne 0} \vert
\langle\phi_n\vert {\rm J} \vert\phi_0
\rangle\vert^2\delta(\omega-(E_n-E_0^{M})),
\label{kubo}
\end{equation}
where ${\rm J}$ is the current operator. 
The finite frequency part (second term) can
be straightforwardly calculated by means of the Lanczos 
algorithm. The coefficient of the zero-frequency delta function contribution
$D$ is usually called the Drude weight and can be evaluated 
by combining the above with the
well known sumrule for the total conductivity in terms of the
kinetic energy $\langle H_t\rangle$,
\begin{equation}
\label{sumhol}
\int_0^{\infty}\sigma(\omega)d\omega = 
-{\pi e^2\over 2}\langle H_t\rangle.
\end{equation}

\begin{figure}
\centerline{\psfig{bbllx=600pt,bblly=100pt,bburx=60pt,bbury=700pt,%
figure=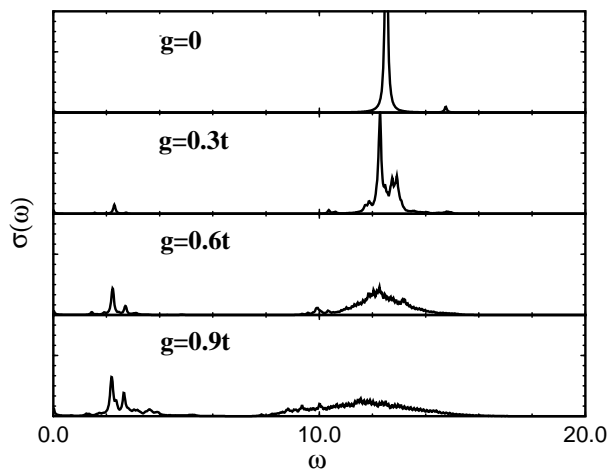,width=70mm,angle=-90}}
\caption{Finite frequency optical conductivity for $U=10t$ and $\omega_0/t = 0.4$,
for various values of $g$ and $n=4/5$}  
\label{sigma}
\end{figure}

In Fig. \ref{sigma} we show the behavior of the finite-frequency part
of the dynamical conductivity for the same parameters as in Fig.
\ref{akw}. Again the most apparent effect of
the coupling with the phonons is a substantial broadening of the
high-energy excitations.  A second remarkable feature is represented
by the marked increase of spectral weight at low frequencies.
As in the case of the single-particle spectral density,
this occurs because the possibility of
phonon excitations accompanying electron particle-hole excitations
allows for a large number of 
intraband transitions (i.e. between states with the same number
of doubly occupied sites). This effect is made particularly apparent 
by the comparison with the $g=0$ case (panel (a) in Fig. \ref{sigma}),
where the small number of sites and of holes strongly suppresses
the weight at frequencies of order $t$ \cite{notasig}. 

As far as the zero-frequency contribution is concerned, 
although quasiparticle transport is absent in $d=1$, there is still
coherent charge (holon) transport. At T=0 this leads to the presence
of a $\delta$-like Drude conductivity at zero frequency. 
It is noticeable that this ideal conductance behavior
persists at finite values of $g$ at least at $T=0$ despite the
non-integrable character of the model with $g\ne 0$.

\begin{figure}
\centerline{\psfig{bbllx=600pt,bblly=100pt,bburx=60pt,bbury=700pt,%
figure=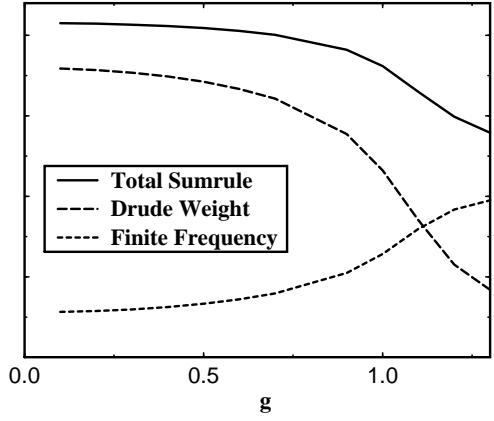,width=70mm,angle=-90}}
\caption{Integrated optical conductivity (Total Sumrule), Drude weight
and finite frequency weight as functions of $g$ for $U=10t$, $\omega_0/t =0.4$ 
and $n=4/5$}  
\label{weights}
\end{figure}

The suppression of coherent charge transport via phonon-dressing
of the holon excitations leads to a strong decrease of
the Drude spectral weight in the dynamical conductivity. 
As shown in Fig. \ref{weights}, the  Drude weight
related to absorption from coherent excitations at $\omega=0$ decreases
slowly in the weak-coupling regime ($\lambda \lesssim 1$)
whereas it is strongly suppressed for larger e-ph coupling
when the polaronic crossover sets in.
In the latter case the rapid decrease is consistent with
the usual exponential decrease of the charge-carrier band-width
$v \to v \exp{-\alpha^2}$.
The only difference is now that in $d=1$, $v$ no longer
is to be interpreted as the Fermi velocity of quasiparticles, but is related
to the holon dispersion.
On the other hand, by increasing $g$, spectral weight is transferred
to higher energies, as indicated by the increase of the finite frequency
weight in Fig. \ref{weights}. 

Although we cannot easily access the very strong coupling regime,
we also checked in a few cases that
for stronger couplings than those reported in the figure, 
this incoherent high-energy part of the absorption also decreases,
but more slowly, according to the expected $1/g$  behavior.
The overall decrease of the total absorption is a natural consequence of
the sum rule (\ref{sumhol}) relating the frequency integrated dynamical
conductivity to the average kinetic energy.

\section{CONCLUSIONS}

In this work, we have investigated the electron-polaron crossover in the
many-particle case. We mainly considered the adiabatic regime, in which 
the condition for polaron formation is modified with respect to the
single-particle case. The main result is that a clear criterion has been
identified determining the strength of e-ph coupling leading to
polaron formation. Remarkably, a simple physical interpretation of the
criterion is still possible in the many-particle case provided that
the single electron energy is replaced by the many-body purely electronic
energy and the magnetic energy contribution is considered in the
polaronic phase. 

The effects of the intermediate polaronic regime
have also been investigated on the dynamical properties.
In particular, both in the single-particle spectral density and in the
optical conductivity, we find that the electron-to-polaron 
crossover leads to a broadening of the high energy ($\omega \sim U$)
features. At the same time the increase of momentum conserving processes
due to the mixed electron-multiphonon character of the excitations
substantially increases the weight of intraband ($\omega \sim t$)
processes.

The above scenario was investigated by numerical exact diagonalization
of a 5-site cluster. This leads to several limitations.
First of all the presence of finite size effects in such a small
system forces the analysis to be limited to short-range models.
Even the analysis of models with a
nearest-neighbor  intersite e-ph coupling like the Su-Schrieffer-Heeger model
lead to substantially larger finite-size effects. 
Another quite interesting problem, which we could not address
within our small-cluster analysis is related to the competition
with other phonon-driven instabilities like  charge-density-waves (CDW),
phase separation or incommensurate stripe formation.
Even in the case of these latter two instabilities, which are not
Fermi-surface instabilities, the small size of our system does
not allow for a systematic comparison with the polaron formation
within our numeric analysis.
On the other hand, some information can be gained
by comparing our results with  different analyses. 
In particular as far as phase separation and (with the inclusion of
long-range Coulomb interactions) stripe formation,
the analyses of Refs.\onlinecite{becca,GC} indicate that they occur for
e-ph couplings that are smaller than those 
leading to polaron formation. 
The approximate character of these analyses, however, still leaves
room for futher work in this direction.
As far as commensurate CDW formation is concerned,
the recent density-matrix renormalization group
analysis of Ref.\onlinecite{bursill} in the infinite-U Hubbard-Holstein model
at $n=1/4$ indicates that the CDW
 instability competes strongly with polaron
formation. However, in the adiabatic regime the CDW and the polaron 
instability occur nearby in the phase diagram  
(whereas polarons are favored in the 
antiadiabatic regime), so it is likely that the polaronic 
instability  will win for
fillings far from 1/4, which are less favorable for CDW.

Part of this work has been carried out with the financial
support of the INFM, PRA HTSC 1996.
The Authors would like to acknowledge Prof. S. Ciuchi, C. Castellani, 
C. Di Castro
and D. Feinberg for many fruitful discussions. 

\end{multicols}

\begin{table}
 \begin{center}
  \begin{tabular}{|c|c|c|c|c|c|c|}
$n$ & U   & ${\cal E}_{el}$  & ${\cal E}_{el}^J$ & $\lambda_c$ & 
$\tilde\lambda_c$ & 
$\tilde\lambda_c'$ \\
   \hline
1/6 & $\infty$ & -2.00000 & 0.       & 1.187 & 0.989 & 1.340 \\
1/5 & $\infty$ & -2.00000 & 0.       & 1.228 & 0.983 & 1.308 \\
2/6 & $\infty$ & -3.46410 & 0.       & 1.069 & 0.823 & 1.183 \\
2/5 & $\infty$ & -3.23607 & 0.       & 1.076 & 0.798 & 1.143 \\
2/5 & 10       & -3.48806 & -0.14253 & 1.269 & 0.910 & 1.187  \\
2/5 & 20       & -3.35407 & -0.06861 & 1.136 & 0.826 & 1.143  \\
2/5 & $J=0.4$  & -3.35407 & -0.12572 & 1.182 & 0.883 & 1.186  \\
2/5 & $J=0.2$  & -3.29317 & -0.05896 & 1.102 & 0.821 & 1.152  \\
3/5 & 10       & -3.71681 & -0.30940 & 1.099 & 0.793 & 1.087  \\
3/5 & 20       & -3.46485 & -0.14124 & 1.079 & 0.788 & 1.109  \\
3/5 & $J=0.4$  & -3.50730 & -0.28743 & 1.002 & 0.767 & 1.109  \\
3/5 & $J=0.2$  & -3.36766 & -0.13559 & 1.040 & 0.781 & 1.124  \\
4/5 & 10       & -3.22487 & -0.92167 & 1.463 & 1.012 & 1.181  \\
4/5 & 20       & -2.60681 & -0.45633 & 1.343 & 0.999 & 1.228  \\
4/5 & $J=0.4$  & -2.90651 & -0.91216 & 1.201 & 0.975 & 1.204  \\
4/5 & $J=0.2$  & -2.45174 & -0.45336 & 1.215 & 0.978 & 1.246
   \end{tabular}
  \end{center}
 \caption{Filling (first column) and interaction (second column)
dependencies of the total purely electronic energy (third column),
of the purely magnetic energy (fourth column), 
of $\lambda_c$ found with the maximum derivative
of $\langle n_{ph}(g)\rangle$ (fifth column), 
and of the $\tilde\lambda_c$'s
calculated {\it without}($\tilde\lambda_c$) and {\it with} 
($\tilde\lambda_c'$the ${\cal E}_{1/\lambda}$ correction in
${\cal E}_{pol}$ for $\omega_0/t =0.2$}
\end{table}   

\begin{multicols}{2}

\end{multicols}


\begin{references}
\bibitem{bcnote} We choose periodic or antiperiodic
boundary conditions in order 
to have a closed shell ground-state for every filling.
\bibitem{notacriterio} Obviously a sligthly different value for the
crossover e-ph coupling could be obtained by calculating different
physical quantities like, e.g., the (electron-density)--(ionic shift)
correlation function, but this would not affect our general
conclusions.
\bibitem{notacohe}
In this regard we notice in passing, that the e-ph coupling
would be a natural probe to identify the different (propagating
vs incoherent) character of the excitations of electronic
systems in more than one dimension.
\bibitem{notasig} We explicitly checked that, in the purely
electronic case with five
sites and four electrons, no transitions are permitted at 
frequencies of order $t$. This feature is not generic, and changes
with system size and/or filling, but this does
emphasize the role of phonons in opening new momentum-conserving
channels available for optical transitions.
\end{references}
\end{document}